\documentclass[aps,pre,twocolumn]{revtex4-1}

\usepackage{amsmath,amssymb,mathrsfs,bm}
\usepackage{color,graphicx,subfigure}

\newcommand{\bA}{\mathbf{A}}
\newcommand{\ba}{\mathbf{a}}

\newcommand{\bb}{\mathbf{b}}
\newcommand{\bBe}{\mathbf{B}_{\mathrm{e}}}
\newcommand{\bBeC}{\bBe^{\,\bigtriangledown}}
\newcommand{\bbD}{\mathbb{D}}

\newcommand{\bD}{\mathbf{D}}
\newcommand{\be}{\mathbf{e}}
\newcommand{\bF}{\mathbf{F}}
\newcommand{\bFe}{\mathbf{F}_{\!\mathsf{e}}}
\newcommand{\bG}{\mathbf{G}}
\newcommand{\bh}{\mathbf{h}}
\newcommand{\bhs}{\mathbf{h}_{*}}
\newcommand{\bH}{\mathbf{H}}
\newcommand{\bI}{\mathbf{I}}

\newcommand{\bPsi}{\bm{\Psi}}
\newcommand{\bPsis}{\bm{\Psi}_{*}}
\newcommand{\bPsiC}{\bPsi^{\,\bigtriangledown}}
\newcommand{\bPsisC}{\bPsis^{\,\bigtriangledown}}

\newcommand{\bn}{\mathbf{n}}
\newcommand{\bns}{\mathbf{n}_{*}}
\newcommand{\bnsp}{\dot{\mathbf{n}}_{*}}
\newcommand{\bno}{\mathbf{\mathring{n}}}
\newcommand{\bnso}{\mathbf{\mathring{n}}_{*}}

\newcommand{\bt}{\mathbf{t}}
\newcommand{\bg}{\mathbf{g}}
\newcommand{\bmm}{\mathbf{m}}
\newcommand{\bnu}{\bm{\nu}}

\newcommand{\bQ}{\mathbf{Q}}
\newcommand{\bT}{\mathbf{T}}
\newcommand{\bW}{\mathbf{W}}

\newcommand{\bv}{\mathbf{v}}

\newcommand{\Pt}{\mathcal{P}_t}
\newcommand{\gm}{\lambda}
\newcommand{\taux}{\tau_{*}}
\newcommand{\taudef}{\tau_{\text{def}}}
\newcommand{\ang}{\Theta}

\newcommand{\eps}{\varepsilon}

\newcommand{\si}{\sigma}
\newcommand{\sio}{\sigma_{0}}

\newcommand{\Wext}{W^{(\text{ext})}}
\newcommand{\FF}{\mathcal{F}}

\newcommand{\Ldot}[1]{\overset{\bm{.}}{#1}}
\newcommand{\gammap}{\Ldot{\gamma}}
\newcommand{\thetap}{\Ldot{\theta}}

\DeclareMathOperator{\tp}{\otimes}

\DeclareMathOperator{\tr}{tr}

\DeclareMathOperator{\divr}{div}

\newcommand{\D}[2]{\frac{\partial #1}{\partial #2}}

\begin{document}

\title{Tumbling in nematic polymers and liquid crystals}
%

\author{Stefano S.\ Turzi}
\affiliation{\mbox{Dipartimento di Matematica, Politecnico di Milano, Piazza Leonardo da Vinci 32, 20133 Milano, Italy}}

\date{\today}

\begin{abstract}
Most, but not all, liquid crystals tend to align when subject to shear flow, while most nematic polymeric liquid 
crystals undergo a tumbling instability, where the director rotate with the flow. The reasons of this instability remain 
elusive, as it is possible to find similar molecules exhibiting opposite behaviors. We propose a theory suitable for describing a wide range of material behaviors, ranging form nematic elastomers to nematic polymers and nematic liquid 
crystals, where the physical origins of tumbling emerge clearly. There are two possible ways to relax the 
internal stress in a nematic material. The first is the reorganization of the polymer network, the second is the alignment of the network natural axis with respect to the principal direction of the effective strain.  Tumbling occurs whenever the 
second mechanism is less efficient than the first and this is measured by a single material parameter $\xi$. 
Furthermore, we provide a justification of the experimental fact that at high temperatures, in an isotropic phase, only 
flow alignment is observed and no tumbling is possible, even in polymers.
\end{abstract}

\maketitle

\section{Introduction}
\label{sec:introduction}

Nematic liquid crystals and nematic polymers either undergo shear aligning or tumbling when subjected to a simple 
shear flow. Shear aligning is characterized by a director dynamics that evolves monotonically 
to a fix orientation at steady state. By contrast, tumbling occurs when the director undergo periodic oscillations in 
its orientation with a period that is inversely proportional to the shear rate. The classical Ericksen-Leslie model of 
nematic liquid crystals predicts that the tendency of a liquid crystal to either tumble or flow align is controlled by 
the sign of the ratio of two viscosity coefficients: $\alpha_3/\alpha_2$. Positive values lead to flow aligning and 
negative values cause tumbling.

However, the different sign of $\alpha_2$ and $\alpha_3$ has no clear chemical or physical interpretation. Thus, a 
number of possible explanations for such a dramatic difference in the flow dynamics has been proposed in the 
literature over the years. It had long been thought that prolate nematogens always align with the flow, then it was 
discovered that some nematic liquid crystals undergo a tumbling instability in part of their nematic range. By 
contrast, most side-chain polymer liquid crystals show tumbling most of the times \cite{00Kamien}.

Furthermore, for some compounds, these different behaviors are unexpected, since their molecular structure and their phase diagrams are very 
similar \cite{95Larson,13Fatriansyah,03zakharov}. For example, while MBBA and 5CB always flow align in their nematic phase, other closely related molecules, respectively HBAB and 8CB, undergo a transition from flow alignment to tumbling when the temperature is decreased below a 
given threshold \cite{95Larson,13Fatriansyah}. 

This 
instability has been associated with smectic fluctuations in the nematic 
phase \cite{99quijada}, with strong side-to-side molecular aggregation \cite{72Gahwiller,95Archer} and, in other 
theories \cite{Chan2005}, the rotational friction, the order parameter
strength and molecular form factors play a key role. Many theories of nematic liquid crystals fail to predict the 
transition from flow aligning to tumbling behavior. Some retain the transition, but the interpretation they provide 
for the tumbling parameter is not widely accepted.

One successful strategy to relate the viscous response of NLCs to the effective mesoscopic features of the 
microscopic constituents, is to derive a kinetic theory of NLCs. This was originally developed by Kuzuu and Doi 
\cite{81Doi,83KuzuuDoi,84KuzuuDoi} and then extended by Osipov and Terentjev \cite{89Osipov,89Osipov2,Chan2005} and 
Larson \cite{90Larson,95Larson,95Archer}. In general, kinetic theories require specific assumptions on 
the intermolecular potential, and this choice particularly affects the antisymmetric part of the Cauchy stress tensor, 
which is responsible for the director rotations. Furthermore, these theories usually include higher order moments of the 
orientational distribution function and some closure approximations are usually necessary to make the theory more 
tractable. However, the most obvious closure approximations imply that NLCs always exhibit flow aligning 
\cite{90Larson}, thus more sophisticated closure approximations are often used to be able include the tumbling effect. 
Finally, kinetic theories neglects the analysis of the translational molecular degrees of freedom, which give a 
fundamental contribution to the Newtonian viscosity (the Leslie coefficient $\alpha_4$). 

In this paper, we propose an alternative route to include far from equilibrium effects, such as tumbling, into a continuum theory. Namely, we develop a ``mixed'' theory where the macroscopic degrees of freedom are treated classically, but the microscopic degrees of freedom are taken into account in a coarse grained way by introducing material reorganization and relaxation. The advantage of this approach is its simplicity, the guiding principles being material symmetry and irreversible thermodynamics. While in this case not all the microscopic details can be accounted for (like in kinetic theories, however), nonetheless we get a better insight into the microscopic mechanisms underlying tumbling phenomena.

The paper is organized as follows: the theory is described in Sec.\ref{sec:dynamics} and \ref{sec:governing_eqns}. 
In Sec.\ref{sec:Leslie} and \ref{sec:tumbling} we obtain some consequences of the theory, in the approximation of 
fast relaxation times. Specifically, we derive how the Leslie coefficient depend on the model parameters and discuss 
the tumbling phenomenon. The conclusions are drawn in Sec.\ref{sec:conclusion}. Finally, some mathematical details on 
the derivation are reported in the appendices.

\section{Natural polymer network}
\label{sec:dynamics}

In our previous papers \cite{14bdt,16bdt,16tur} we have shown how nematic elastomers (NEs), nematic liquid crystals 
(NLCs) and nematic polymers (NPs) can be described, at the continuum level, by the same theory. \emph{A-posteriori}, 
this is not surprising since they share the basic features of a continuum theory, namely, material symmetry and 
compatibility with thermo-mechanics principles.  

If we model nematic elastomers as rubbery networks with an aligned 
uniaxial anisotropy of their polymer strands and with a coupling to the nematic mesogenic units, the transition from a 
elastic response of NEs to a fluidlike behavior of NLCs is obtained by allowing the polymer network to reorganize. 
Hence, we consider a transient polymer network, where cross-links can break under stress at some rate and reforms 
in an unstressed state, so that the network undergoes a plastic deformation to reach a natural state with zero stress, 
a state that we call \emph{natural (or relaxed) polymer network}. 
However, at short time-scales, when the cross-links are not broken, the material is elastic. In general, when the 
cross-linking rate is much higher than the breakage rate the network
can be regarded as ``cross-linked'', or elastic.  When the two rates are comparable the system undergoes a 
plastic flow under stress and when cross-linking rate is much lower than the breakage rate, the system quickly relaxes 
to a natural state and it behaves like a viscous fluid. Of course, in real nematic liquid crystals the network is not 
physical, but it is only an idealization. Its transient nature mimics the rearrangement of the position of the nematic 
molecules that typically takes place in fluids.

Usually the positional order of the cross-links or of the nematic molecules, at each instant of time, is only known via 
some averaged quantities. A standard simplification in this respect is that the second moment tensor is sufficient to 
describe the positional distribution of the molecules. Hence, in analogy with nematic elastomer theory \cite{Warner}, 
we define a \emph{shape tensor} $\bPsis$ as the chain step-length tensor of the \emph{relaxed} network (or, depending 
on interpretation as the normalized covariance tensor of the one-particle probability density of the position of the 
molecules), 
\begin{equation}
\bPsis(\rho,\bns) = a(\rho)^2 (\bns \tp \bns) + a(\rho)^{-1}\big(\bI - \bns \tp \bns \big),
\label{eq:shape_tensor}
\end{equation}
where $\rho$ is the density, $a(\rho)$ is a shape parameter that gives the amount of spontaneous elongation along the 
main axis $\bns$. When 
$a(\rho) = 1$, the centers of mass distribution is isotropic, while for $a(\rho) > 1$ ($<1$) it is prolate 
(respectively, oblate) in the direction of $\bns$. The normalization condition corresponds to the requirement 
$\det(\bPsis)=1$, since we are only concerned with the \emph{anisotropy} of the molecular distribution. Our definition 
of $\bPsis$ mimics the definition of step-length-tensor that is used in nematic elastomer theory to describe the 
anisotropic polymer ordering and represents the spontaneous stretch of the material.

In standard nematic elastomer theory, and in our previous works \cite{14bdt,16bdt}, it is assumed that the relaxed 
 network main axis, $\bns$, is directed along the nematic director $\bn$, at each instant of time. In so doing, the 
director is taken to describe, at the same time, the preferred orientation of the molecules and the relative distance of 
their centers of mass at equilibrium (or the direction of the natural strain in the network). This assumption is valid 
for most NLCs and for NEs and leads to interesting consequences such as the connection of the Leslie 
coefficients with the elastic features and the relaxation times of the material, the dependence of viscosity 
coefficients on frequency and the viscoelastic response of the material. Furthermore, a new Parodi-like relation is 
identified for NLCs which seems to be in good agreement with molecular simulations and in fairly good agreement with 
experiments.

\begin{figure}
\includegraphics[width = 0.4\textwidth]{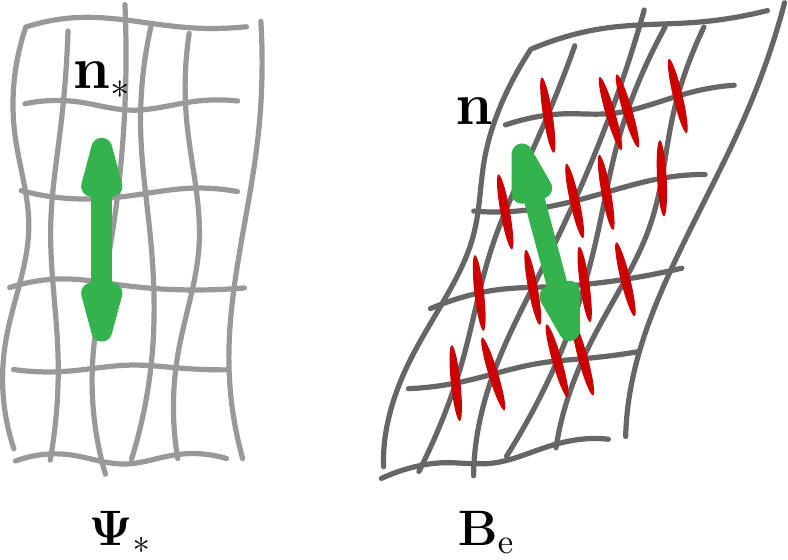}
\caption{Schematic representation of the natural network, with main axis $\bns$, whose anisotropy is described via the shape tensor $\bPsis$ (left). For comparison, on the right hand side we show the schematic representation of the deformed actual transient network, with effective strain tensor $\bBe$. The material is stress-free when $\bBe$ coincides with $\bPsis$. The director $\bn$ describes the average direction of the molecules and is an additional degree of freedom, but it is energetically coupled with the axis $\bns$. }
\label{fig:TransientNetwork}
\end{figure}

However, the assumption of an instantaneous relaxation of $\bns$ to the director $\bn$ leads to a flow-aligning 
director field and prevents tumbling instability, which is observed in most nematic polymers and in some liquid 
crystals. Therefore, a key ingredient for the presence of tumbling seems to be the distinction 
between $\bPsis$ and the elastonematic-coupling tensor defined as
\begin{equation}
\bPsi(\rho,\bn) = a(\rho)^2 (\bn \tp \bn) + a(\rho)^{-1}\big(\bI - \bn \tp \bn \big),
\label{eq:elastonematic}
\end{equation}
the only difference with respect to \eqref{eq:shape_tensor} being the substitution of $\bns$ with $\bn$ (see Fig. 
\eqref{fig:TransientNetwork}). Our 
description differs at this point from the standard elasticity of nematic elastomers where there is a direct coupling 
between the director field and the polymer network. Instead of a single shape-tensor, 
here we introduce \emph{two} closely related tensors, $\bPsis$ and $\bPsi$, and the coupling between their axes, 
$\bns$ and $\bn$, is described by an energetic term that favors the alignment of $\bns$ with $\bn$. As we shall see, 
this introduces an additional governing equation and a corresponding characteristic time.

To account for the transient nature of the network (or material reorganization, in the case of NLCs), we split the 
deformation gradient, $\bF$, in elastic part ($\bFe$) and relaxing part ($\bG$), and define
\begin{align}
\bFe & = \bF\bG^{-1}\, , \quad \bH = (\bG^{T}\bG)^{-1} \, , \\
\bBe & = \bFe\bFe^{T} =\bF\bG^{-1} \bG^{-T}\bF^{T} = \bF \bH \bF^{T},
\end{align}
where $\bBe$ is the effective left Cauchy-Green deformation tensor, and $\bH$ is the inverse relaxing strain tensor.

Furthermore, we posit the following free energy density per unit mass
\begin{align}
\si(\rho,\bBe,\bns,& \bn,\nabla\bn) = \sio(\rho) + \tfrac{1}{2}\mu_1(\rho) \Big(\tr \big(\bPsis^{-1}\bBe - \bI \big) 
\notag \\
& - \log\det\big(\bPsis^{-1}\bBe \big)\Big) \notag \\
& + \tfrac{1}{2}\mu_2(\rho) \tr \big(\bPsis^{-1}\bPsi - \bI \big)
+ \sigma_{\text{Fr}}(\bn, \nabla\bn).
\label{eq:DensitaEnergiaPsi}
\end{align}
The first term $\sio(\rho)$ penalizes volume changes, it is assumed to be large and not depending on 
material relaxation. The second term represents a neo-Hookean 
energy where the natural (zero stress) deformation is described by 
$\bPsis$ and is, therefore, transversely isotropic in the direction of $\bns$. This free-energy term only depends 
on the effective tensor $\bBe$ which is allowed to relax to its natural state $\bPsis$. The bulk modulus 
associated with the elastic response is $\rho \mu_1(\rho)$. However $\bns$ (and hence $\bPsis$) is not fixed, but can 
rotate in order to align with the director $\bn$. 
This contribution is encoded in the third term which penalizes any deviation of $\bns$ from the director $\bn$. The relative importance of this term with respect to the second one is 
determined by the ratio of their elastic moduli: $\mu_2/\mu_1$. Finally, we consider a Frank elastic-potential 
$\sigma_{\text{Fr}}(\bn, \nabla\bn)$ that favors the alignment of 
nematic molecules and whose prototype is $\sigma_{\text{Fr}}(\bn, \nabla\bn) = k \,|\nabla \bn|^2$. For more complex 
scenarios in this respect, the reader is advised to consult Ref.\cite{95dgpr}.

Intuitively, the dynamics of $\bns$ is governed by two independent 
contributions: on the one hand, $\bns$ is coupled to the effective macroscopic 
deformation $\bBe$, so that $\bns$ tends to align with a principal direction of the effective strain; on the other 
hand $\bns$ tends to coincide with the director $\bn$. The director $\bn$ is an additional degree of freedom, so if there was no Frank potential, it would always be 
 favorable to make the director align with the major axis of $\bBe$, and take $\bns=\bn$. However, in the 
presence of director elastic energy this configuration could have a high energy cost due to possible director 
distortions, so that an intermediate configuration could be preferable. In such a case it is possible that $\bns$ and 
$\bn$ do not coincide.

The minimum of the free energy, which is reached at equilibrium, is yielded by
\begin{equation}
\bBe = \bPsis, \qquad \bPsis = \bPsi. 
\label{eq:equilibrium}
\end{equation}
In dynamics, far from equilibrium, identities \eqref{eq:equilibrium} do not generally hold. However, when $\mu_1 \ll 
\mu_2$ the third term in \eqref{eq:DensitaEnergiaPsi} dominates the second, so that it would be too energetically 
expensive to have two different shape tensors for long times. If the relaxation dynamics is 
sufficiently fast (in a sense to be specified later), we can assume to leading order $\bPsis = \bPsi$ and 
thus recover our previous theory by considering only the second term. When $\mu_1$ and $\mu_2$ are comparable, or the 
relaxation dynamics is slow, we need to keep two separate shape tensors and study the dynamics that brings $\bPsis$ to 
evolve towards $\bPsi$, or equivalently, $\bns$ in the direction of $\bn$. This dynamics is governed by an additional 
characteristic time, $\taux$, introduced below. 

\section{Governing equations}
\label{sec:governing_eqns}
Here, we simply state the main equations of the model. The full derivation is given in Appendix 
\ref{app:derivation}. More details on the physical meaning of some terms may also be gathered from 
\cite{14bdt,16bdt,15tur,16tur,17tur}.

There are two type of governing equations. The first set of equations comprises balance laws that do 
not imply dissipation of energy. In our case these are the equations for the velocity field $\bv$ and the director $\bn$
\begin{align}
\rho \dot{\bv} = \bb + \divr\bT, \qquad \bn \times (\bg - \bh) = 0,
\label{eq:balance}
\end{align}
where an overdot indicates the material time derivative. The boundary conditions are 
\begin{align}
\bt_{(\bnu)} = \bT\bnu, \qquad 
\bn \times \bmm_{(\bnu)} = \bn \times \left(\rho\D{\si}{\nabla\bn}\right) \bnu.
\end{align}
In the above equations, $\bnu$ is the outer unit normal to the surface, $\bt_{(\bnu)}$ and $\bmm_{(\bnu)}$ are the 
surface tractions and the surface couples, $\bT$ is Cauchy stress tensor, $\bb$ is the external body force, $\bh$ is 
the nematic molecular field and $\bg$ is the external field acting on the director (e.g. a magnetic field) which will be 
be set to zero in the following for simplicity. As shown in the appendix, $\bT$ and $\bh$ are defined as
\begin{align}
\bT & := \rho\D{\si}{\bF}\bF^{T} - \rho(\nabla \bn)^T\D{\si}{\nabla\bn} \notag \\
& \qquad + \frac{1}{2} (\bhs \tp \bns - \bns \tp \bhs) \label{eq:Tdef} \\
\bh & := \rho\D{\si}{\bn} - \divr\left(\rho\D{\si}{\nabla\bn} \right), \label{eq:hdef}
\end{align}
where the molecular field $\bhs$ associated to $\bns$ is 
\begin{equation}
\bhs := \rho\D{\si}{\bns}. \label{eq:hsdef}  
\end{equation}

The second type of equations are associated with irreversible processes and follow from linear irreversible 
thermodynamics principles. These equations describe how the effective strain tensor, $\bBe$, and the main axis of the 
natural polymer network, $\bns$, evolve
\begin{align}
\bbD (\bBeC) & = -\D{\sigma}{\bBe} \label{eq:evoluzione_Be} \\
\gm (\bns \times \bnso) & = - \bns \times \bhs. \label{eq:evoluzione_ns} 
\end{align}
The kinematics of the material reorganization is described by the upper-convected time derivative $\bBeC$ and the 
corotational derivative of $\bns$, defined as 
\begin{align}
\bBeC & := (\bBe)\Ldot{\phantom{I}} - (\nabla\bv) \,\bBe - \bBe \,(\nabla\bv)^T. \label{eq:bBeC}\\
\bnso & := \bnsp - \bW \bns, 
\label{eq:corotational}
\end{align}
where $\bW = \tfrac{1}{2}\big(\nabla \bv - (\nabla \bv)^T\big)$ is the spin tensor. The tensor $\bbD$ 
is a fourth-rank tensor which is compatible with the uniaxial symmetry about $\bns$, has the major symmetries and is 
positive definite \cite{16tur}, and $\gm$ is a positive material parameter. These phenomenological quantities 
contains the characteristic times of material reorganization and specify what are the possible different modes of 
relaxation and how fast these relaxation modes drag the system to equilibrium. 

For our purposes, it suffices to say that $\bbD$ comprises four relaxation times: $\tau_1$, $\tau_2$, $\tau_3$ and 
$\tau_4$ (see \cite{16tur} for details), while $\gm$ leads to the introduction of a fifth relaxation time $\taux$, 
defined in the next Section. Specifically, $\tau_1$ measures the relaxation time of the \emph{pure} shearing modes in a 
plane through $\bns$ (i.e., a stretching that makes a 45$^\circ$ angle with $\bns$, and does not involve rotations). By 
contrast $\tau_2$ is associated with pure sharing modes that happen in the plane orthogonal to $\bns$.  

\section{Leslie coefficients}
\label{sec:Leslie}
In this section we derive the explicit expressions for the Cauchy stress tensor \eqref{eq:Tdef}, the molecular fields 
\eqref{eq:hdef}, \eqref{eq:hsdef} and the relaxation equations \eqref{eq:evoluzione_Be},  
\eqref{eq:evoluzione_ns} when the free energy $\si$ is given as in Eq.\eqref{eq:DensitaEnergiaPsi}. This will allow us 
to simplify our model for fast relaxation times and thus to give a physical interpretation of the Leslie coefficients 
in terms of our model parameters. For ease of reading part of the calculations are reported in Appendix 
\ref{app:Leslie}.

A little algebra, allows us to rearrange the Cauchy stress tensor, as given in \eqref{eq:Tdef} (or \eqref{eq:Tdef2}), 
in the following form
\begin{align}
\bT & = - p\,\bI + \rho\mu_1 \Big( \bPsis^{-1}\bBe - \bI \Big) \notag \\
& -\frac{\gm}{2} (\bnso \tp \bns - \bns \tp \bnso)
-\rho (\nabla\bn)^T \D{\si_{\text{Fr}}}{\nabla\bn} , \label{eq:Tdef3}
\end{align}
with $p := \rho^2 \D{\si}{\rho}$ a pressure-like function. 
The material reorganization is governed by \eqref{eq:evoluzione_Be}, which takes the form
\begin{align}
\hat{\bbD}(\bBeC) - \bBe^{-1} + \bPsis^{-1} = 0
\label{eq:evoluzione_Be_2}
\end{align}
where $\hat{\bbD} = 2\bbD/\mu_1$ is simply proportional to $\bbD$ but it has been rescaled in order to have dimensions 
of time. The director equation (\ref{eq:balance}b) and the relaxation equation \eqref{eq:evoluzione_ns} are found to be 
(see Appendix \ref{app:Leslie} for details)
\begin{align}
\divr\left(\rho\D{\si_{\text{Fr}}}{\nabla\bn} \right) & 
+ \rho \mu_2 \frac{(a^3-1)^2}{a^3}(\bns \cdot \bn)(\bn \times \bns) = 0 \label{eq:bilancio_n_2} \\
\taux (\bns \times \bnso) & = \frac{\mu_1}{\mu_2} \, \frac{a^3}{a^3-1}\, (\bns \times \bBe\bns) \notag \\
& + (\bns \cdot \bn)(\bns \times \bn)
\label{eq:evoluzione_ns_2}
\end{align}
where 
\begin{equation}
\taux = \frac{a^3}{(a^3-1)^2}\frac{\gm}{\rho \mu_2} 
\label{eq:taux}
\end{equation}
is proportional to the parameter $\lambda$ and can be taken as the characteristic time associated with the 
reorientation of $\bns$. At the end of Sec.\ref{sec:governing_eqns} we have seen that when the material undergoes a pure 
shear strain in a plane through $\bns$, the effective strain tensor relaxes to the natural state with a characteristic 
time $\tau_1$. An alternative mechanism to relax the internal stress is to rotate the unit cell of the natural network 
(and its main axis $\bns$) in order to conform to a general superimposed deformation or in response to a mismatch with 
the director field $\bn$, and this happens with a characteristic time $\taux$. 

It is easy to see that $\tau_1$ and $\taux$ are independent times and indeed can be very different. Let us consider a 
nematic elastomer. Its rubbery network is elastic and does not reorganize, so that $\tau_1$ can be considered 
infinite. By contrast, $\taux$ is finite and is interpreted as the time that the director $\bn$ takes to coincide with 
the principal direction of the superimposed strain (for NEs $\bn = \bns$ by assumption).

It is interesting to observe that, according to Eq. \eqref{eq:bilancio_n_2}, when the director field is 
homogeneous, i.e., $|\nabla \bn|=0$, $\bns$ is either parallel 
or orthogonal to $\bn$. This can also be seen from the energy density. Whenever the Frank potential vanishes in 
\eqref{eq:DensitaEnergiaPsi}, every possible configuration of minimum energy satisfies $\bPsis = \bPsi$, or, in 
other terms, $\bns = \bn$. Any deviation from $\bns = \bn$ costs some energy and this excess energy is ultimately due 
to distortions in the director field. 

When $\mu_1 \ll \mu_2$ and we assume that $\tau_i$ ($i=1,2,3,4$) and $\taux$ are much smaller that the characteristic 
times associated with the deformation, measured by $\taudef = \max\{1/|\nabla \bv|\}$, $\bBe$ and $\bns$ are just a 
small corrections of their equilibrium values $\bPsis$ and $\bn$. Eq. \eqref{eq:evoluzione_Be_2} then yields the 
approximation of $\bBe$ to first order
\begin{equation}
\bBe \approx \bPsis - \bPsis\,\hat{\bbD}(\bPsisC) \bPsis.
\end{equation}
This approximation is suitable for the description of a fluidlike behavior and, therefore, it is appropriate for NLCs 
and possibly for some nematic polymers (when viscoelastic effects are not important), but it is not applicable to the 
other possible extreme of the model, namely, nematic elastomers. 

To obtain the approximation of the stress tensor \eqref{eq:Tdef3} to first order, it is sufficient to consider only 
the leading term of \eqref{eq:evoluzione_ns_2}. To leading order, $\bns \approx \bn$, so that \eqref{eq:Tdef3} becomes
\begin{align}
\bT & = - p \bI - \rho\mu_1 \hat{\bbD}(\bPsiC) \bPsi \notag \\
& - \frac{(a^3-1)^2}{2a^3}\rho\mu_2 \taux (\bno \tp \bn - \bn \tp \bno) \notag \\
& - \rho (\nabla\bn)^T \D{\si_{\text{Fr}}}{\nabla\bn} . \label{eq:Tdef4}
\end{align}
If we now compare \eqref{eq:Tdef4} with the classical expression of the Cauchy stress tensor, as given by the 
compressible Ericksen-Leslie theory, 
\begin{equation}
\begin{split}
\bT & = -p\bI + \alpha_1 (\bn\cdot \bD\bn)(\bn\tp\bn) + \alpha_2 (\bno \tp  \bn) + \alpha_3 (\bn \tp  \bno) \\
& + \alpha_4 \bD + \alpha_5 (\bD\bn \tp \bn) + \alpha_6 (\bn \tp \bD\bn) \notag \\
& + \alpha_7 \big((\tr \bD ) (\bn\tp\bn) + (\bn\cdot \bD\bn)\bI \big) + \alpha_8 (\tr \bD ) \bI,
\end{split}
\label{eq:TLeslie}
\end{equation}
and use the explicit expression for $\hat{\bbD}$ as given in Ref.\cite{16tur}, we obtain that the Leslie coefficients 
in terms of our model parameters are
\begin{subequations}
\begin{align}
\alpha_1 & = \rho\mu_1(\rho) \Big(\tau_2 - \frac{\left(a(\rho )^3+1\right)^2}{a(\rho )^3}\tau_1
+ 3 \tau_3 (\cos\ang)^2 \notag \\
& + 3\tau_4 (\sin\ang)^2 \Big), \\
\alpha_2 & = - \rho\mu_1(\rho) \tau_1 \left(a(\rho )^3-1\right)
- \rho \mu_2(\rho) \taux \frac{(a(\rho)^3 - 1)^2}{2 a(\rho)^3} , \\
\alpha_3 & =  - \rho\mu_1(\rho) \tau_1 \left(1 - a(\rho)^{-3}\right)
+ \rho \mu_2(\rho) \taux \frac{(a(\rho)^3 - 1)^2}{2 a(\rho)^3},  \\
\alpha_4 & = 2 \rho\mu_1(\rho) \tau_2,  \\
\alpha_5 & = \rho \mu_1(\rho) \Big(\left(1+a(\rho)^3\right)\tau_1 - 2 \tau_2 \Big), \\
\alpha_6 & = \rho \mu_1(\rho)  \Big(\left(1+a(\rho)^{-3}\right)\tau_1 - 2 \tau_2 \Big),
%
\end{align}
\label{eq:alphas2}
\end{subequations}
where $\ang$ is an additional parameter that appears in the definition of $\bbD$ and affects $\alpha_1$ but plays no 
role in what follows. It is also possible to find the bulk viscosity coefficients $\alpha_7$ and $\alpha_8$, but these 
are not particularly relevant for the purposes of the present paper, and we omit them for brevity.

It is interesting to observe that, in agreement with experiments, $\alpha_2$ is always negative for rod-like LCs 
($a(\rho) > 1$) as it is obtained as the sum of two negative terms. By contrast, $\alpha_3$ can be either negative or 
positive, the latter case leading to a tumbling behavior. Vice versa, for disk-like molecules $\alpha_3$ is always 
positive while $\alpha_2$ can be positive (flow alignment) or negative (tumbling).

The Parodi relation is automatically satisfied along with a 
second identity \cite{16bdt}
\begin{equation}
\alpha_6-\alpha_5 = \alpha_2 + \alpha_3, \qquad 
\frac{\alpha_4 + \alpha_5}{\alpha_4+\alpha_6} = \frac{\alpha_2 + \gm/2}{\alpha_3 - \gm/2} = a(\rho)^3, 
\end{equation}
where, we recall, $\gm = \rho \mu_2 \taux\,(a^3-1)^2/a^3$.

\section{Tumbling parameter}
\label{sec:tumbling}
\begin{figure}
\begin{center}
\includegraphics[width=0.4\textwidth]{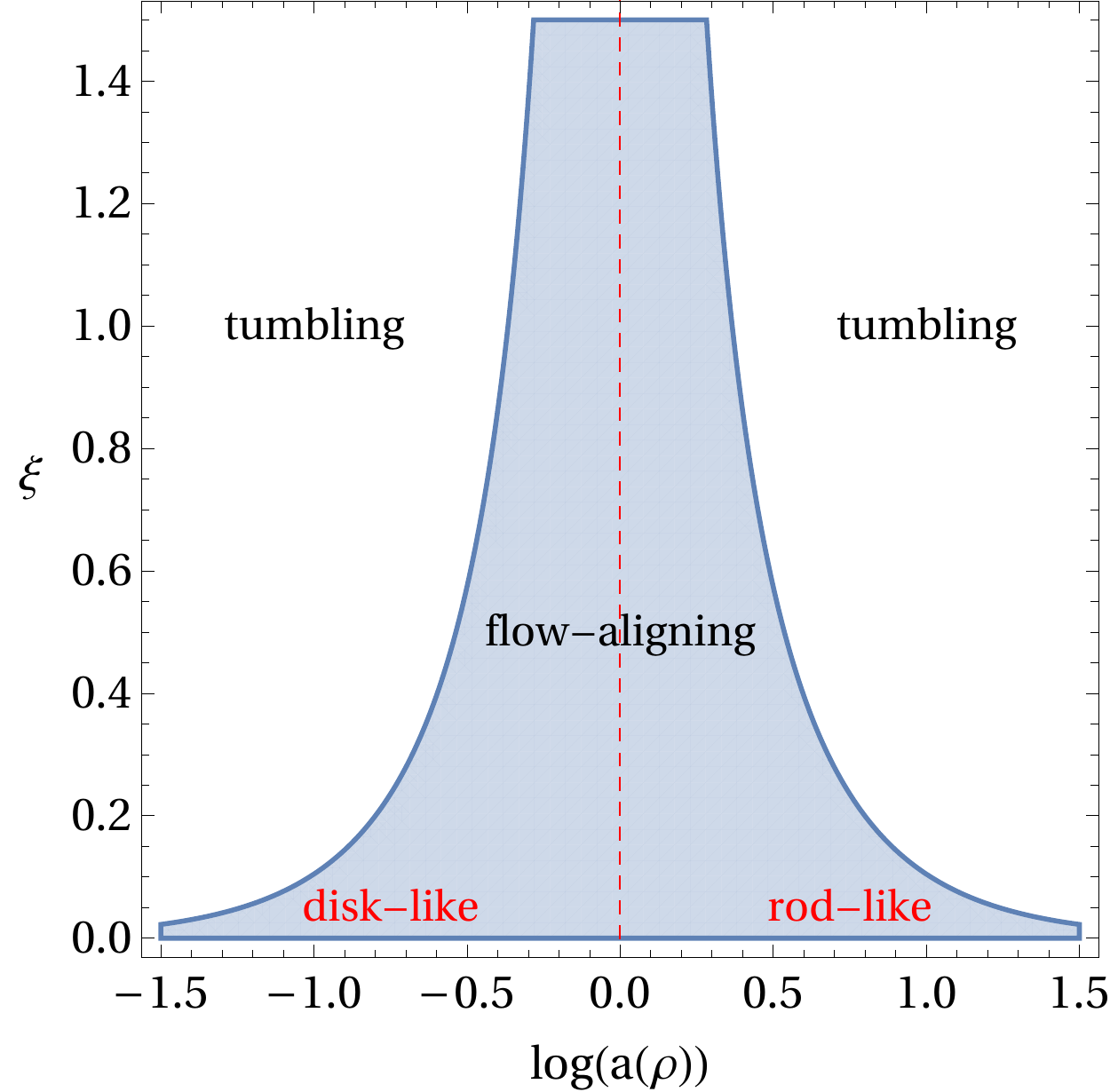} 
\end{center}
\caption{Flow-aligning (blue) and tumbling (white) regions, as deduced from \eqref{eq:tumblingregion}, as a function of 
the model parameters $a(\rho)$ and $\xi$. The region with $a(\rho) > 1$ corresponds to prolate shape tensors,  
associated with rod-like molecules. Disk-like molecules correspond to $a(\rho)<1$. Tumbling ceases to exist in the 
isotropic phase, where the shape tensors are spherical ($a(\rho)=1$).}
\label{fig:tumbling_diagram}
\end{figure}

In terms of Leslie coefficients, it is known that a tumbling instability arises whenever $\alpha_3/\alpha_2 < 0$ 
\cite{95dgpr,Chan2005}, a condition that, after the substitution of \eqref{eq:alphas2}, reads
\begin{equation}
\frac{\alpha_3}{\alpha_2} = \frac{1}{a(\rho)^3} \left(\frac{2 - \xi (a(\rho)^3-1)}{2 - \xi (a(\rho )^{-3}-1)}\right) < 
0,
\label{eq:tumbling_condition}
\end{equation}
where we have defined the key ratio
\begin{equation}
\xi = \frac{\mu_2 \, \taux}{\mu_1 \, \tau_1}.
\label{eq:csi}
\end{equation}
In fact, the flow alignment angle is known to be
\begin{align}
\tan \theta = \sqrt{\frac{\alpha_3}{\alpha_2}}, 
\end{align}
so that alignment is only possible when $\alpha_2$ and $\alpha_3$ have the same sign (both positive or negative). If 
$\xi \approx 0$, so that the natural network axis is free to reorient with the flow, only flow-alignment is 
possible. By simplifying \eqref{eq:tumbling_condition}, we get the following condition for tumbling behavior
\begin{equation}
\begin{cases}
\displaystyle \xi > \frac{2}{a(\rho)^{-3}-1} 
& \text{ if } \quad 0 < a(\rho) < 1, \\[5mm]
\displaystyle \xi > \frac{2}{a(\rho)^3-1} 
& \text{ if } \quad a(\rho) > 1. 
\end{cases}
\label{eq:tumblingregion}
\end{equation}
In either case tumbling occurs when the ratio $\frac{\mu_2 \, \taux}{\mu_1 \, \tau_1}$ is larger than a given 
threshold, which depends on the anisotropy of the shape tensor (i.e., on the aspect ratio $a(\rho)$). This threshold 
goes to infinity in the isotropic case $a(\rho)=1$ and decreases for increasing anisotropy of the shape tensor (see 
Figure \ref{fig:tumbling_diagram}). Even if our theory do not depend on temperature, it is reasonable to expect that in 
the isotropic phase the shape tensor become spherical, i.e., $a(\rho)=1$. Hence, we get a clear explanation of 
why tumbling is enhanced by the presence of orientation order, and it is suppressed in the isotropic phase, even 
for nematic polymers.

We have seen that $\xi$ represents the ratio between two possible effects: relaxation by alignment of the 
natural network axis with the flow, and strain relaxation by material reorganization. As is clear from 
Fig.\ref{fig:tumbling_diagram}, if the first mechanism prevails, the material 
flow-aligns, while if the second is more efficient, the director tumbles (when sufficiently far from the isotropic 
phase). This fact is more pronounced for long-chains or flat-disks, i.e., it increases with molecular anisotropy. 

This interpretation is in agreement with previous experiments and with some earlier theoretical claims, where tumbling 
was explained in terms of strong side-to-side molecular association \cite{72Gahwiller,95Archer,94Gu}. In particular, 
the authors of Ref.\cite{94Gu} observed experimentally that ``addition of a side-chain LCP [liquid crystal
polymer] to flow-aligning 5CB induces a director-tumbling response, whereas dissolution of a main-chain
LCP in director-tumbling 8CB induces a flow-aligning response.'' This agrees with our interpretation in that strong 
side-chain association may hinder natural network rotations while it does not have much influence on the material 
reorganization by network sub-cell sliding.

A more direct way to reach the same conclusions, that does not make use of the Leslie coefficients, is to study the 
evolution of the natural network main axis using Eq. \eqref{eq:evoluzione_ns_2}. In the absence of nematic distortions, 
from the balance equation \eqref{eq:bilancio_n_2}, we get that $\bn=\bns$. Hence, in the limit $\mu_2\taux \ll 
\mu_1\tau_1$, 
Eq.\eqref{eq:evoluzione_ns_2} to leading order  reads $\bns \times \bBe\bns = 0$, i.e., $\bns$ aligns with a principal 
direction of the effective strain (which is fixed in the fast relaxation approximation discussed above). In this case 
$\bns$ is constant and aligns with the flow. On the other 
hand, if $\mu_2\taux \gg \mu_1\tau_1$, to leading order we have $\bns \times \bnso = 0$. Since $\bnso$ is orthogonal to 
$\bns$, this implies that $\bnso$ must vanish. In such a case, we do not obtain a steady solution, but a rotating field: 
$ \bnsp = \bW \bns$.

More precisely, let us consider a simple shear 
flow in a semi-infinite medium, under the usual assumption of 
fast relaxation approximation ($\mu_2 \gg \mu_1$ 
and $\taux$, $\tau_i \ll \taudef$). The velocity field is written as $\bv = \gammap \, y \,\be_x$, where $\gammap$ 
is the shear rate and $\be_x$ is a unit vector along the $x$-axis. Within our approximation, 
$\bns$ is just a small perturbation of its equilibrium value $\bn$ and we posit
\begin{equation}
\eps = \mu_1/\mu_2, \qquad  \bns = \bn + \bn_1, \qquad \bn_1 = O(\eps),
\end{equation}
where $\bn$ is taken to be homogeneous over the whole sample (i.e., $|\nabla \bn |=0$), $\bn_1 \cdot \bn=0$ and $\bn_1$ 
measures the difference between $\bns$ and $\bn$ in a dynamic situation. 
The relaxation times are much smaller than $\taudef$, so that the product of either $\taux$ or 
$\tau_1$ with a time-derivative (i.e., $\bno$ or $\bD\bn$) is taken to be $O(\eps)$.
For a homogeneous director field, the balance equation \eqref{eq:bilancio_n_2} implies that $\bn_1$ 
vanishes to first order, so that we can assume $\bns \approx \bn$. Hence, Eq.\eqref{eq:evoluzione_ns_2} reads
\begin{align}
\taux (\bn \times \bno) & = - \frac{\mu_1}{\mu_2} \, \frac{a^3}{a^3-1}\, 
[\bn \times \bPsi\,\hat{\bbD}(\bPsiC) \bPsi\bn\,].
\label{eq:XXX}
\end{align}
The codeformational derivative of $\bPsi$ can be explicitly calculated in terms of the imposed macroscopic flow so that 
 Eq.\eqref{eq:XXX} simplifies to 
\begin{align}
\xi (\bn \times \bno) = - \bn \times \left(\bno - \frac{a^3 + 1}{a^3-1}\bD\bn \right).
\label{eq:XXX2}
\end{align}
We now introduce the tilt angle $\theta$ such that $\bn = \sin\theta(t) \, \be_x + \cos\theta(t)\, \be_y$. After a 
little algebra \eqref{eq:XXX2} can be written as
\begin{align}
(1+\xi)\,\thetap = \frac{\gammap}{2} \left(1+\xi + \frac{a^3 + 1}{a^3-1} \cos(2\theta) \right).
\label{eq:XXX3}
\end{align}
When the liquid crystal aligns with the macroscopic flow, the angle $\theta$ is constant, so that $\thetap=0$. 
Thus, stationary solutions are only possible if 
\begin{equation}
\frac{|a^3-1|}{a^3+1} \leq \frac{1}{1+\xi}, 
\label{eq:tumbling_2}
\end{equation}
a condition that, after some simplification, is shown to coincide with the flow-aligning condition
$\alpha_3/\alpha_2 \geq 0$, as derived from Leslie coefficients \eqref{eq:alphas2}, and is complementary to the 
tumbling condition \eqref{eq:tumblingregion}. When the condition 
\eqref{eq:tumbling_2} is not met (or, equivalently, the complementary condition \eqref{eq:tumblingregion} holds), we 
can still 
solve \eqref{eq:XXX3} in this regime to obtain the periodic oscillations of the director, i.e, the functional 
dependence of $\theta$ over time.

\section{Conclusions}
\label{sec:conclusion}

The mechanisms underlying tumbling instability are subtle and there is no widely accepted explanation for the physical origins of this phenomenon. We find that the distinction between the nematic director and the principal axis of the natural polymer network is the key feature to observe the crossover between flow-aligning and tumbling behaviors.

This distinction allows the material, when it undergoes a shearing deformation, to relax the internal stress in two distinct ways. The first
is the internal reorganization of the polymer network cross-links, the second is the rotation of the natural polymer network main axis to align with the principal direction of the effective strain. Both these mechanisms reduce the internal stress, but tumbling occurs whenever the first mechanism prevails over the second. 

In agreement with previous claims, this explanation suggests that tumbling is due to a strong side-to-side molecular association, either by electrostatic interactions or by steric interaction (for example in long flexible polymer chains). In our model a single material parameter $\xi$, defined as the ratio $\mu_2 \taux / \mu_1 \tau_1$, describes the relative importance of one mechanism over the other. Furthermore, we show that in the isotropic phase only flow-aligning is possible and that tumbling is enhanced by strong molecular anisotropy. 

However, in order to fully study the tumbling dependence on the degree of order and the temperature effects, it is necessary to construct a theory that includes the nematic ordering tensor $\bQ$. It is expected that the resulting theory in this case be highly non-trivial and its analysis is postponed to a following paper. 

Another important reason for introducing the tensor $\bQ$ in our model, is that tumbling typically generates defects \cite{96Mather}, indicating that the system could no longer be regarded as a monodomain \cite{13Fatriansyah}. It is observed in \cite{96Mather} that the defect structures, or textures, in a 8CB sample depends on the shear history of the sample. In particular, the texture depends not only on the rotation speed, but also on the rate at which the rotation speed is increased from zero. This is in agreement with the viscoelastic nature of our model and the consequent interpretation of tumbling in terms of relaxation processes. By contrast, the Ericksen-Leslie model, having frequency-independent viscosity coefficients, cannot reproduce different material behaviors or aligning features for different shear rates or shear histories.


\begin{acknowledgments} I would like to thank Antonio Di Carlo and Paolo Biscari for their valuable and helpful suggestions during the planning and development of this research.

\end{acknowledgments}

\appendix
\section{Derivation of the model}
\label{app:derivation}
In this section we derive the governing equations. The second principle of thermodynamics requires that, for any 
isothermal process, for any portion $\Pt$ of the body at all times, the dissipation (rate of entropy production) be 
greater or equal than zero \cite{GurtinFried}
\begin{align}
\mathcal{D} := W^{(\text{ext})} - \dot{K} - \dot{\FF} \geq 0 , 
\end{align}
where $\Wext$ is the power expended by the external forces, $\dot{K}$ is the rate of change of the kinetic energy, 
$\dot{\FF}$ is the rate of change of the free energy, and the dissipation $\mathcal{D}$ is a positive quantity that 
represents the energy loss due to irreversible process. Here, an overdot indicates the material time derivative. More 
precisely, we define
\begin{align}
W^{(\text{ext})} & = \int_{\Pt} \bb\cdot \bv \,\,dv + \int_{\partial\Pt} \bt_{(\bnu)}\cdot \bv \,\,da \notag\\
& + \int_{\Pt} \bg \cdot \dot{\bn} \,\,dv + \int_{\partial\Pt} \bmm_{(\bnu)} \cdot \dot{\bn} \,\,da, \\
K + \FF & = \int_{\Pt} \left(\frac{1}{2}\rho \bv^2 + \rho \si(\rho,\bBe,\bns,\bn,\nabla\bn) \right) \,dv, \\
\mathcal{D} & = \int_{\Pt} \xi \,\,dv, \qquad \xi \geq 0.
\end{align}
The unit vector $\bnu$ is the external unit normal to the boundary 
$\partial\Pt$; $\bb$ is the external body force,  $\bt_{(\bnu)}$ is the external 
traction on the bounding surface $\partial\Pt$. The vector fields $\bg$ and 
$\bmm_{(\bnu)}$ are the external generalized forces conjugate to the 
microstructure: $\bn \times \bg$ is usually interpreted as ``external body 
moment'' and $\bn \times \bmm_{(\bnu)}$ is interpreted as ``surface moment per 
unit area'' (the couple stress vector).

The material time-derivative of $\FF$ is
\begin{align}
\dot{\FF} 
& = \int_{\Pt} \Big(\rho\D{\si}{\bF}\cdot \Ldot{\bF} + \rho\D{\si}{\bH}\cdot \Ldot{\bH} \notag \\
& + \rho \D{\si}{\bn}\cdot \Ldot{\bn} + \rho \D{\si}{\bns}\cdot \bnsp 
+ \rho \D{\si}{\nabla\bn}\cdot (\nabla\bn)\dot{\phantom{i}} \Big) dv.
\label{eq:Fdot1}
\end{align}
If we introduce the (frame-indifferent) upper-convected time-derivative $\bBeC$, as given in Eq.\eqref{eq:bBeC}
and use the identities
\begin{align}
\dot{\bF} & = (\nabla\bv) \bF, \\
\frac{D}{Dt} (\nabla\bn) & = \nabla \Ldot{\bn} - (\nabla \bn)(\nabla \bv), \\
\D{\si}{\bBe} & = \bF^{-T}\D{\si}{\bH}\bF^{-1}, \\
\bBeC & = \bF\Ldot{\bH}\bF^{T},
\end{align}
Eq. \eqref{eq:Fdot1} simplifies to
\begin{align}
\dot{\FF} & = \int_{\Pt} \Big(\rho\D{\si}{\bF}\bF^{T}\cdot \nabla \bv + \rho\D{\si}{\bBe}\cdot \bBeC
+ \rho \D{\si}{\bn}\cdot \Ldot{\bn} \notag \\
& + \rho \D{\si}{\bns}\cdot \bnsp + \rho \D{\si}{\nabla\bn}\cdot (\nabla\bn)\dot{\phantom{i}} \Big) dv, \notag \\
& = \int_{\Pt} \left(\rho\D{\si}{\bF}\bF^{T} - \rho(\nabla \bn)^T\D{\si}{\nabla\bn} \right) \cdot 
\nabla \bv \,\,dv \notag \\
& + \int_{\Pt} \left(\rho\D{\si}{\bn}\cdot \Ldot{\bn} + \rho\D{\si}{\nabla\bn} \cdot \nabla\Ldot{\bn}\right) \,\, dv 
\notag \\
& + \int_{\Pt} \left(\rho\D{\si}{\bns} \cdot \bnsp \right) \,\, dv
+ \int_{\Pt}\rho\D{\si}{\bBe}\cdot \bBeC\,\, dv.
\end{align}
We now define the molecular fields $\bh$ and $\bhs$, as in Eqs. \eqref{eq:hdef}, \eqref{eq:hsdef} so that 
we rewrite the second integral as
\begin{align*}
\int_{\Pt} \Big(\rho\D{\si}{\bn} & +\rho\D{\si}{\nabla\bn} \cdot \nabla\Ldot{\bn}\Big) \,\, dv \notag \\
& = \int_{\partial\Pt} \left(\rho\D{\si}{\nabla\bn}\right) \bnu \cdot \Ldot{\bn}\,\, da  
+ \int_{\Pt} \bh \cdot\Ldot{\bn} \,\, dv.
\end{align*}
Furthermore, since we assume that a relaxed network realignment gives a positive dissipation, we have 
to recast the third integral in terms of frame-indifferent fields (no dissipation is 
associated with a rigid body rotation of the whole body). Hence, we write
\begin{align}
\int_{\Pt} \bhs \cdot \bnsp \,\, dv 
 = \int_{\Pt} \bhs \cdot \bnso \,\, dv + \int_{\Pt} \bhs \cdot \bW\bns \,\, dv \notag \\
 = \int_{\Pt} \bhs \cdot \bnso \,\, dv + \int_{\Pt} \frac{1}{2} (\bhs \tp \bns - \bns \tp \bhs) \cdot \nabla \bv \,\, 
dv.
\end{align}
The last term is paired with $\nabla \bv$ so that it represents a contribution to the Cauchy 
stress tensor, defined as in Eq.\eqref{eq:Tdef} and repeated here for convenience
\begin{align}
\bT & := \rho\D{\si}{\bF}\bF^{T} - \rho(\nabla \bn)^T\D{\si}{\nabla\bn}
+ \frac{1}{2} (\bhs \tp \bns - \bns \tp \bhs). \label{eq:Tdef2}
\end{align}
Therefore, the final expression for the rate of change of the free energy is
\begin{align}
\dot{\FF} 
& = \int_{\Pt} \bT \cdot \nabla \bv \,\,dv 
+ \int_{\partial\Pt} \left(\rho\D{\si}{\nabla\bn}\right) \bnu \cdot \Ldot{\bn}\,\, da \notag \\
& + \int_{\Pt} \bh \cdot \Ldot{\bn} \,\, dv + \int_{\Pt} \bhs \cdot \bnso \,\, dv
+ \int_{\Pt}\rho\D{\si}{\bBe}\cdot \bBeC\,\, dv \notag \\
& = \int_{\partial\Pt} \bT\bnu \cdot \bv \,\,da - \int_{\Pt} \divr\bT \cdot \bv \,\,dv \notag \\
& + \int_{\partial\Pt} \left(\rho\D{\si}{\nabla\bn}\right) \bnu \cdot \Ldot{\bn}\,\, da 
+ \int_{\Pt} \bh \cdot \Ldot{\bn} \,\, dv \notag \\
& + \int_{\Pt} \bhs \cdot \bnso \,\, dv
+ \int_{\Pt}\rho\D{\si}{\bBe}\cdot \bBeC\,\, dv,
\end{align}
and the dissipation is then written as
\begin{align}
\mathcal{D} & = W^{(\text{ext})} - \dot{K}-\dot{\FF} \notag \\
& = \int_{\Pt} \left(\bb - \rho \dot{\bv} + \divr\bT \right) \cdot \bv \,\,dv 
+ \int_{\partial\Pt} \left(\bt_{(\bnu)} - \bT\bnu \right) \cdot \bv \,\,da \notag \\
& + \int_{\Pt} \left(\bg - \bh \right) \cdot \Ldot{\bn} \,\,dv 
+ \int_{\partial\Pt} \left(\bmm_{(\bnu)}  - \left(\rho\D{\si}{\nabla\bn}\right) \bnu \right) \cdot \dot{\bn} \,\,da 
\notag \\
& -\int_{\Pt} \bhs \cdot \bnso \,\, dv - \int_{\Pt}\rho\D{\si}{\bBe}\cdot \bBeC\,\, dv.
\label{eq:dissipazioneIntegrale}
\end{align}

By assumption, a positive dissipation is associated to material reorganization and only the last two integrals can 
contribute to the irreversible processes (i.e., can have a positive dissipation). Thus, the contribution from the first 
integrals must vanish and we have the equations (we recall that $\bn \cdot \Ldot{\bn} = 0$) for the deformation field 
$\bv$ and the director field $\bn$, as given in Eqs.\eqref{eq:balance}.
The dissipation then simplifies to 
\begin{align}
\mathcal{D} & =  - \int_{\Pt} \left(\bhs \cdot \bnso + \rho\D{\si}{\bBe}\cdot \bBeC \right) \, dv.
\end{align}
A simple choice that satisfies $\mathcal{D} \geq 0$ at all times and is consistent with standard linear irreversible 
thermodynamics \cite{95dgpr,DeGroot} is to take the fluxes proportional to forces. In 
so doing, we arrive at Eqs.\eqref{eq:evoluzione_Be} and \eqref{eq:evoluzione_ns}.
Furthermore, we assume Onsager reciprocal relations, so that the proportionality coefficient $\bbD$ is a fourth-rank 
tensor which is compatible with the uniaxial symmetry about $\bns$, has the major 
symmetries and is positive definite (i.e., such that $\bbD(\bA) \cdot \bA >0, \, \forall \bA \neq 
0$ and symmetric), while $\gm$ is a positive material parameter. Indeed, it can be shown that only one coefficient 
$\gm$ is necessary in \eqref{eq:evoluzione_ns} for symmetry reasons (see \cite{95dgpr}). Eq. \eqref{eq:evoluzione_ns} 
governs the dynamics that brings $\bns$ towards $\bn$ (or vice-versa) and the parameter $\gm$ contains the 
characteristic time of this relaxation process.

Finally, we note that, if we denote by $\bW_{\ba}$ the skew-symmetric tensor with axial vector $\ba$, we have from 
\eqref{eq:evoluzione_ns}
\begin{align}
\frac{1}{2} (\bhs \tp \bns - \bns \tp \bhs) = \frac{1}{2}\bW_{\bns \times \bhs} \notag\\
= - \frac{\gm}{2}\bW_{\bns \times \bnso}
= -\frac{\gm}{2} (\bnso \tp \bns - \bns \tp \bnso),
\end{align}
so that the Cauchy stress tensor can also be written in a more familiar form as 
\begin{align}
\bT & = \rho\D{\si}{\bF}\bF^{T} - \rho(\nabla \bn)^T\D{\si}{\nabla\bn}
-\frac{\gm}{2} (\bnso \tp \bns - \bns \tp \bnso). \label{eq:Tdef_app}
\end{align}

\section{Derivation of the Leslie coefficients}
\label{app:Leslie}
In order to find the director equation and the the relaxation equations we need to explicitly elaborate on the terms
\begin{align}
\D{\si}{\bBe} & = \frac{1}{2}\mu_1 \Big(\bPsis^{-1} - \bBe^{-1} \Big), \\
\D{\si}{\bn} & = 2 (a^2 - a^{-1}) \D{\si}{\bPsi}\bn \notag \\
& = \mu_2 (a^2 - a^{-1}) \bPsis^{-1} \,\bn, \\
\D{\si}{\bns} & = 2 (a^{-2} - a) \D{\si}{(\bPsis^{-1})}\bns \notag \\
& = (a^{-2} - a) \big(\mu_1\bBe + \mu_2 \bPsi \big)\bns ,\\
\bn \times \bh & = -\rho \mu_2 \frac{(a^3-1)^2}{a^3}(\bns \cdot \bn)(\bn \times \bns) \notag \\
& - \divr\left(\rho\D{\si_{\text{Fr}}}{\nabla\bn} \right), \\
\bns \times \bhs & = \rho\mu_1 (a^{-2} - a) \bns \times \bBe\bns \notag \\
& - \rho \mu_2 \frac{(a^3-1)^2}{a^3}(\bns \cdot 
\bn)(\bns \times \bn).
\end{align}


%

\end{document}